\documentclass{blois}

\def\Journal#1#2#3#4{{#1} {\bf #2}, #3 (#4)}

\def\PRL{\em Phys. Rev. Lett.}
\def\PRD{{\em Phys. Rev.} D}

\def\JCAP{\em J. Cosmol. Astropart. Phys.}

\def\be{\begin{equation}}
\def\ee{\end{equation}}
\def\bea{\begin{eqnarray}}
\def\eea{\end{eqnarray}}

\newcommand{\Photo}{\includegraphics[height=35mm]{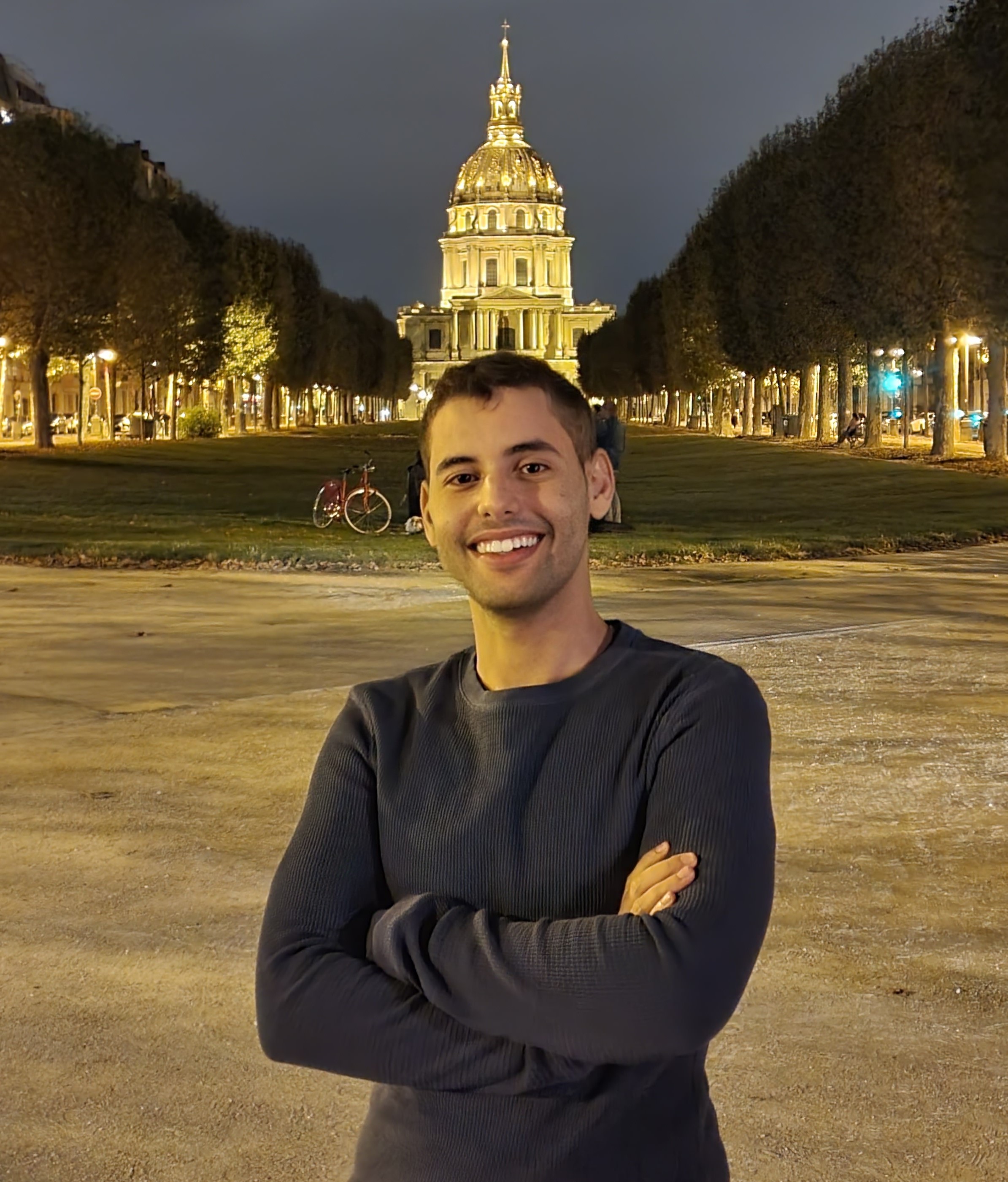}}

\begin{document}
\vspace*{4cm}
\title{STATUS AND FIRST RESULTS FROM XENONNT}

\author{ S.E.M. AHMED MAOULOUD }

\address{LPNHE, Sorbonne Université, CNRS/IN2P3,\\
75005 Paris, France
\bigskip }

\maketitle\abstracts{
XENON is a project for direct dark matter search located in the INFN underground laboratory LNGS. The previous generation of XENON detector, XENON1T, achieved an exposure of 1 ton$\times$year, setting the most stringent limits on the spin-independent scattering cross section of Weakly Interacting Massive Particles (WIMPs) on nucleons for nearly the complete range of WIMP masses above $120$ MeV.
The multi-tonne XENONnT detector is the next step in the evolution of the XENON project. The experiment, aimed at directly detecting WIMPs, utilizes $5.9$~t of instrumented liquid Xenon (LXe). The expected overall background in the detector is based on dedicated screening results of all detector materials. In contrast, improvements in mitigating intrinsic backgrounds from electronic recoil sources allow XENONnT to reduce this background compared to its predecessor. Adding a neutron veto around the XENONnT cryostat also allows for a significant rejection of the neutron background. XENONnT aims at achieving a 20 ton$\times$year exposure over its lifetime.
The talk summarised the main results of XENON1T and the status of XENONnT.}

\section{Introduction}

The XENON dark matter research project operates an underground experiment to detect dark matter particles at the INFN Laboratori Nazionali del Gran Sasso (LNGS) in Italy. To achieve this, the Collaboration seeks signals from rare interactions in the form of nuclear or electronic recoils in a target volume of liquid Xenon. The experiment detects the scintillation and ionization produced when particles interact in the target volume, intending to demonstrate an excess of events, providing the first experimental evidence of direct detection of a dark matter particle. The XENON project has built one of the most promising experiments for directly detecting dark matter.

\begin{figure}[h]
    \centering
    \includegraphics[height=70mm]{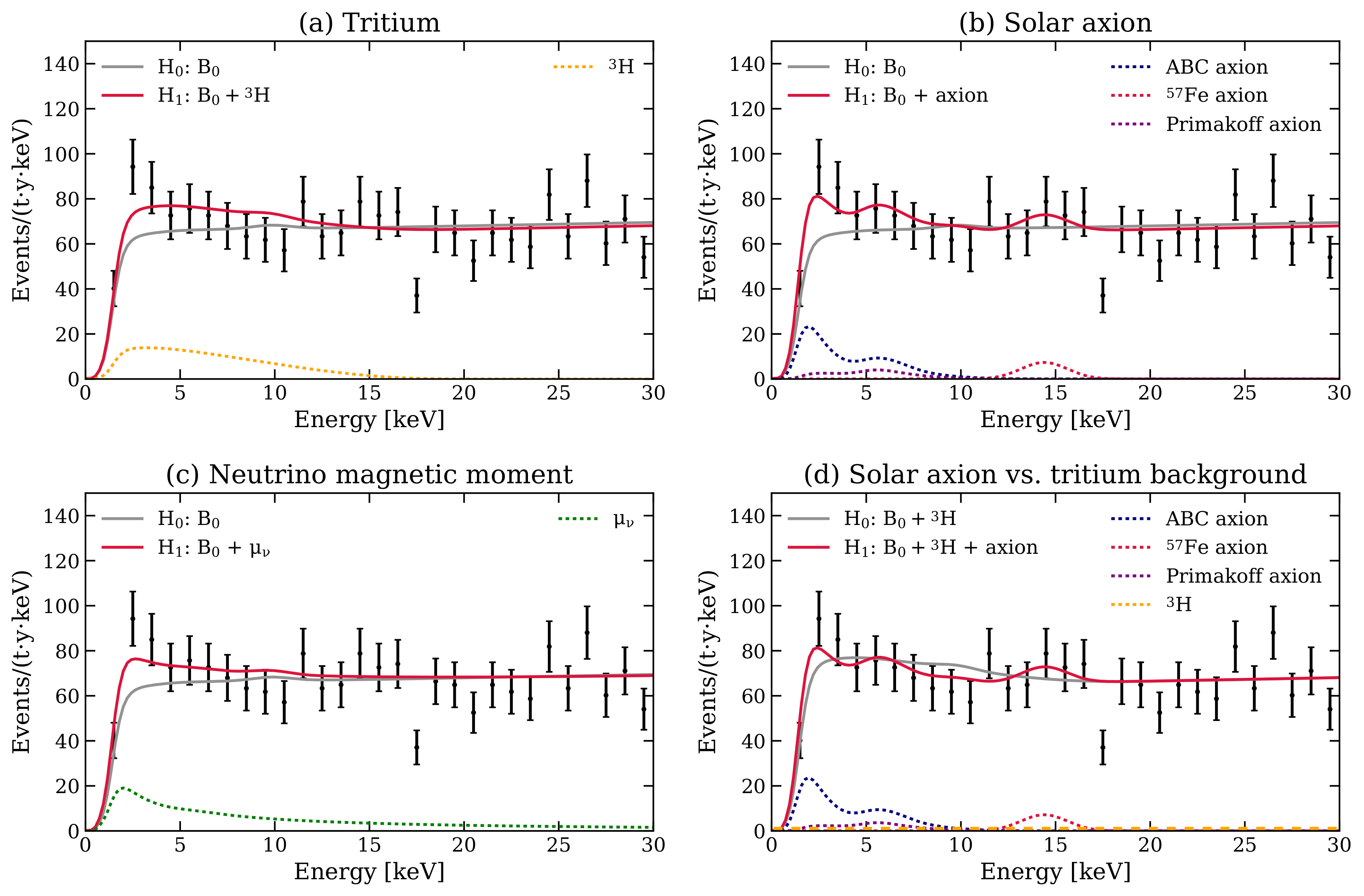}
    \caption{Fits to the data under various hypotheses. For the tritium (a), solar axion (b), and neutrino magnetic moment (c) searches, the null hypothesis is the background model $B_0$, and the alternative hypothesis is $B_0$ plus the respective signal. The null and alternative hypotheses in each scenario are denoted by gray (solid) and red (solid) lines, respectively. Dashed lines illustrate contributions from selected components in each alternative hypothesis. Panel (d) shows the best fit for an additional statistical test on the solar axion hypothesis, where an unconstrained tritium component is included in both null and alternative hypotheses. This tritium component contributes significantly to the null hypothesis. However, its best-fit rate is negligible in the alternative hypothesis, illustrated by the orange dashed line in the same panel. (Figures from ref.~\protect\cite{aprile2020excess})}
    \label{fig:low_energy_fits}
\end{figure}

The previous XENON phase, XENON1T, using $278.8$ days of data collection equivalent to $1$ ton$\times$year exposure, set the most stringent limits on the spin-independent scattering cross section of WIMPs on nucleons for nearly the complete range of dark matter particle masses above $120~$MeV \cite{aprile2018dark,aprile2019light,aprile2019search}. \newline
XENON1T also detected excess low-energy electronic recoil events above the expected background. While this excess may have come from the $\beta$ decay of tritium, which was not previously included in the background model, other possibilities beyond the Standard Model, including solar axions and a magnetic moment of the neutrinos, have been mentioned \cite{aprile2020excess}. With a trace of tritium of the order of $(6.2 \pm 2.0) \cdot 10^{-20}$ mol/mol added to the background model, the anomaly is explained with a significance of $3.2\,\sigma$. In contrast, the significances of the solar axion and neutrino magnetic moment hypotheses are decreased to $2.0\,\sigma$ and $0.9\,\sigma$, respectively, if an unconstrained tritium component is included in the fit (see  \textsc{Figure} \ref{fig:low_energy_fits}).

The signal excess shown in \textsc{Figure} \ref{fig:low_energy_fits} can be further explored with XENONnT, allowing us to study the excess in much more detail if it persists. Preliminary studies suggest a solar axion signal could be differentiated from a tritium background at a $5~\sigma$ level after only a few months of XENONnT data taking.

\subsection{The dual-phase Xenon TPC}
\label{tpc_section}

The XENON Collaboration operates a dual-phase time projection chamber (TPC), which uses a liquid Xenon target with a small gaseous portion above. Two arrays of photomultiplier tubes (PMTs), one at the top of the gas phase and another at the bottom of the liquid detect scintillation and electroluminescence light produced when particles interact in the detector. An electric field is applied vertically across the detector; a second electric field in the gas phase must be large enough to pull electrons out of the liquid phase.

Particle interactions in the liquid Xenon target produce scintillation and ionization. The scintillation light signal is detected within $100$ ns by the PMTs and is called \emph{S1 signal}. The applied electric field prevents the recombination of most electrons produced from the particle interaction in the TPC. These electrons are carried upwards from the liquid phase by the electric field. The stronger electric field applied at the liquid-gas interface pulls the electrons out of the liquid phase. The electric field then accelerates the electrons to the point that they generate a secondary proportional scintillation signal, also known as electroluminescence, as they travel through the thin layer of Xenon gas. This second signal is called the \emph{S2 signal} \cite{aprile2016xenon100}.

\subsection{The XENONnT detector}

The XENONnT main detector is a cylindrical TPC, $148$ cm long and $137$ cm diameter, placed inside a cryostat filled with $8.3$~t of liquid Xenon distributed in two concentric cylinders. The sensitive interior volume (the TPC itself) contains $5.9$~t of Xenon and is separated from the exterior volume by a PTFE cage (polytetrafluoroethylene or Teflon). The PTFE cage has surfaces treated with diamond tools to act as a reflector for Xenon UV light. The outer volume acts as a passive shield by reducing the number of interactions in the inner volume. The veto system reduces the amount of data collected and improves the ability of the detector to reject gamma events as dark matter candidates. This information can be found in the cited source, \cite{aprile2020projected}.

Finally, stainless steel \emph{diving bell} maintains a stable liquid-gas interface between the grid electrodes and the anode. 


\subsection{The neutron veto}
\label{nVeto}

The XENONnT neutron veto aims to reduce the background of radiogenic neutrons by tagging events where the interaction in the TPC coincides with a detected neutron in the neutron veto. 120 Hamamatsu R5912-100-10 8-inch PMTs with a high quantum efficiency (approximately 40\% at $350$ nm) and low radioactivity are placed with reflective panels around the cryostat. The side panels of the neutron veto form an octagonal enclosure of about two meters by three. The PMTs are spread over six equidistant rows. Most of the PMT bodies remain behind the reflective panels to minimize the remaining radioactive background of their materials inside the neutron veto. All reflective surfaces are $1.5$ mm thick expanded PTFE (ePTFE), with light reflectivity greater than 99\% for wavelengths above $280$ nm \cite{aprile2020projected}.

Neutrons that interact within the volume of the TPC can easily pass through the cryostat and evade further detection in LXe. To increase the probability of neutron detection via neutron capture, 0.5\% of Gadolinium sulfate octahydrate will be added to the neutron veto water tank. Therefore, the water around the cryostat moderates neutrons, leaving the TPC volume, typically traveling less than $20$ cm before being thermalized and captured by the Gadolinium with a probability of 91\% (9\% for water) \cite{aprile2020projected}. Following neutron capture by Gadolinium, a cascade of gamma rays with a total energy of about $8$ MeV is generated. In the case of capture by water, a single gamma of $2.2$ MeV is emitted. The energy deposited by gammas in water, mainly by Compton scattering, is converted into electrons and, ultimately, Cherenkov photons.

\section{Projected XENONnT sensitivity}

\begin{figure}[h]
    \centering
    \includegraphics[height=64mm]{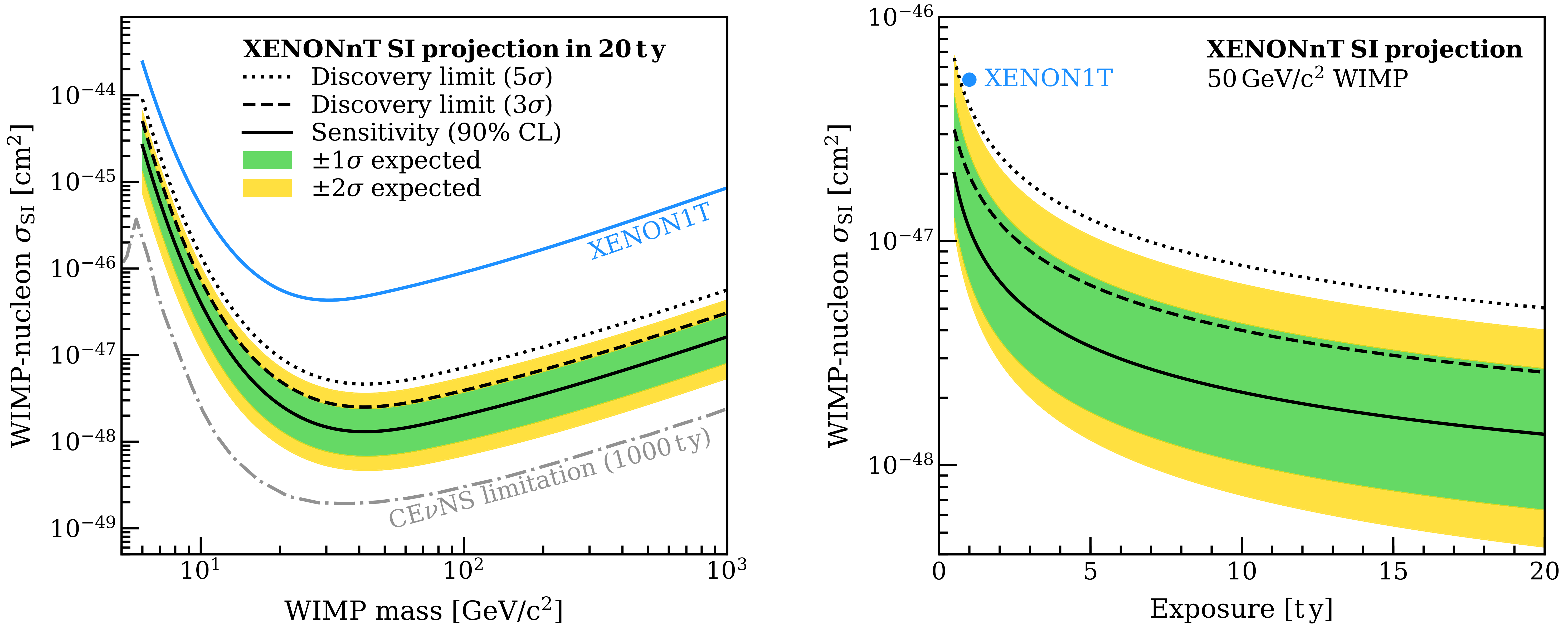}
    \caption{The XENONnT sensitivity and discovery power projections in the search for spin-independent WIMP-nucleon couplings. (Left) Median $90\%$ CL exclusion limit (black solid line) for a 20 ton$\times$year exposure, with the $1\sigma$ (green) and $2\sigma$ (yellow) bands. The current strongest exclusion limit obtained with XENON1T~\protect\cite{aprile2018dark} is shown in blue. The gray dashed-dotted line represents the discovery limit of an idealized LXe-based experiment with CE$\nu$NS as a unique background source and a $1000$ ton$\times$year exposure~\protect\cite{ruppin2014complementarity}. The atmospheric neutrino background would significantly slow down the improvement of the discovery potential with increasing exposure below that line. (Right) Sensitivity as a function of exposure, for searching for a 50\,GeV/c$^2$ WIMP in the assumed $4$~t fiducial mass. The dashed (dotted) black lines in both panels indicate the smallest cross sections at which the experiment would have a $50\%$ chance of observing an excess with significance greater than $3\sigma$ ($5\sigma$). A two-sided profile construction is used to compute the confidence intervals. (Figure from ref. \protect\cite{aprile2020projected})}
    \label{fig:sensitivity_projections}
\end{figure}

\textsc{Figure} ~\ref{fig:sensitivity_projections} shows the expected sensitivity of XENONnT (for the full calculation, see ref. \cite{aprile2020projected}). \textsc{Figure} \ref{fig:sensitivity_projections} (left) expresses the median exclusion limit at 90\% CL on the SI WIMP-nucleon cross section. Thus, with an exposure of 20 ton$\times$year, XENONnT could probe cross sections more than an order of magnitude below the current best limits set by XENON1T~\cite{aprile2018dark}, reaching the highest sensitivity of $1.4\times10^{-48}\,\mathrm{cm}^2$ for a WIMP of 50\,GeV/c$^2$. Projected median discovery levels of XENONnT with significance $3\sigma$ (dashed) and $5\sigma$ (dotted) are shown with sensitivity (solid). The minimal WIMP cross section at which the experiment has a 50\% chance of observing an excess of significance greater than $3\sigma$ ($5\sigma$) is $2.6\times10^{-48}\, \mathrm{cm }^2$ ($5.0\times10^{-48}\,\mathrm{cm}^2$), corresponding to a mass of 50\,GeV/c$^2$. \textsc{Figure} \ref{fig:sensitivity_projections} (right) gives the sensitivity and discovery power for a WIMP of 50\,GeV/c$^2$ as a function of exposure.

\section{Summary}

After the considerable efforts made in designing, assembling, and commissioning the XENONnT detector during arduous periods caused by the COVID-19 pandemic, the XENONnT experiment started taking its first science run (SR0) around the middle of 2021 for about six months. SR0 data are currently being studied, with particular attention given to investigating the excess ER events detected by XENON1T at low energy \cite{aprile2020excess}.
If XENONnT does not detect any dark matter excess during its lifetime, it can nevertheless place some of the best exclusion limits worldwide on the direct search for dark matter. Exclusion limits are helpful for theoretical modeling evaluating past dark matter detection claims and are beneficial information for the next generation of dark matter experiments. \newline
XENONnT has pioneered many new purification and cleaning techniques, which will undoubtedly be used by the next-generation experiment as part of the new XENON-LUX-ZEPLIN-DARWIN (XLZD) consortium \cite{aalbers2016darwin}.

\end{document}